\documentclass[aps,prb,showpacs,twocolumn,preprintnumbers,amsmath,amssymb]{revtex4}
\usepackage{graphicx}
\begin{document}
\title{On the phase diagram of the itinerant helimagnet MnSi: does MnSi become 
quantum critical ?}

\author{Alla E. Petrova}
\affiliation{Institute for High Pressure Physics of Russian
Academy of Sciences, Troitsk, Moscow Region, Russia}
\author{Sergei M. Stishov}
\email{sergei@hppi.troitsk.ru}
\affiliation{Institute for High
Pressure Physics of Russian Academy of Sciences, Troitsk, Moscow
Region, Russia}

\date{\today}

\begin{abstract}
We performed a series of resistivity measurements of a MnSi single crystal at high pressures, created by a piston-cylinder device with a liquid pressure medium.  The form of the resistivity curve at ambient pressure clearly indicates a first order nature of the magnetic phase transition in MnSi. Application of high pressure shows fast degradation of the first order features of the phase transition. The temperature derivative of resistivity demonstrates two notable features of the phase transition that disappear on pressure increasing. They are a sharp peak marking the first order phase transition and a shallow maximum, situated slightly above the critical temperature and pointing to the domain of prominent helical fluctuations. The current experimental data rule out any strong first order phase transition in MnSi at high pressures and low temperatures, which would prevent development of a quantum critical region. On the contrary, there should exist true quantum critical phenomena in MnSi at high pressures because a tiny first order transition, if it survives at high pressures to the lowest temperature, hardly could suppress the entire quantum critical region.

\end{abstract}
\maketitle
Manganese silicide, MnSi, a remarkable example of a helical magnet with the Dzyaloshinski-Moria interaction has attracted much experimental and theoretical interest. But, despite extensive efforts many important features of the temperature-pressure $(T-P)$ phase diagram of MnSi are not well understood. Nevertheless, there is a general belief that a ferromagnetic phase transition inevitably becomes first order as it is tuned toward low temperatures, therefore avoiding development of quantum criticality.  Still, it is not quite clear how the magnetic phase transition in MnSi evolves along the transition line and what are properties of the phase transition in MnSi in the high pressure, low temperature limit. The general situation with the phase transition in MnSi is briefly described below (see review ~\cite{1}).
 A magnetic phase transition of unidentified nature in MnSi at about $30 K$ was reported in Ref.~\cite{2}. Later the magnetic phase of MnSi was identified as a phase with a helical spin order ~\cite{3}. Studies of the heat capacity and thermal expansion of polycrystalline and single crystal samples of MnSi seemingly demonstrated a continuous (second order) nature of the phase transition ~\cite{4}. Since then, the magnetic phase transition in MnSi was generally treated as a second order phase transition, though as was indicated already in 1980 ~\cite{5} the magnetic phase diagram of MnSi provided evidence in favor of a first order transition. 
Subsequently, it was found that applying high pressure drove the phase transition temperature of MnSi toward zero at pressures of about $1.4 GPa$ ~\cite{6}. Later studies at high pressures found a dramatic change in behavior of the ac magnetic susceptibility along the phase transition line.  A peak in the magnetic susceptibility of MnSi at the phase transition degraded into a simple step at $~1.2 GPa$ and $~ 12 K$  that was interpreted as an indication of a tricritical point on the transition line where a second order phase transition became a first order one \cite{7,8}. The "tricritical" idea found theoretical support ~\cite{9} declared a generic nature of first order phase transitions in itinerant ferromagnets at low temperatures that would preclude quantum criticality phenomena. Application of this idea to MnSi led to the concept of a new quantum state of matter (non-Fermi liquid metal without quantum criticality \cite{10}). Later it was shown, however, that disappearance of the peak in magnetic susceptibility of MnSi at the phase transition was most probably caused by non hydrostaticity of a pressure media \cite{11}. 
At a careful study of a good quality single crystal of MnSi \cite{12,13}  sharp peaks (dips) in the temperature derivative of resistivity, the heat capacity and thermal expansion coefficient were observed that led to the conclusion of a weak first order character of the magnetic phase transition in MnSi at ambient pressure. Ultrasound measurements, performed by the same group, confirmed this conclusion \cite{14}. Small angle neutron scattering experiments provided independent proof of a first order phase transition in MnSi \cite{15}.The experiments \cite{12,13,14} also solidly confirmed features (a sharp peak and a shoulder) in the thermal expansion coefficient and sound absorption found in earlier experiments \cite{16,17}. Some earlier heat capacity measurements \cite{18} also revealed a shoulder on the high temperature side of the peak, which became less prominent with increase of RRR (Residual Resistivity Ratio). As a result, the existence of the shoulder could be ascribed to sample imperfection \cite{18}, but we note that long annealing of MnSi, while improving of the RRR, leads to degradation of the phase transition. 
Though mounting evidence \cite{12,13,14}, proving a first order transition in MnSi at ambient pressure, initially was ignored in favor of a tricritical scenario, in time a first order character of the magnetic phase transition in MnSi at ambient pressure finally has been recognized \cite{15,19, 20, 21}. 
So the current situation is the following. At ambient pressure MnSi experiences the first order magnetic transition near 29 K with a tiny volume discontinuity $-\Delta V/V\approx 3\times10^{-6}$ Ref.\cite{12}. At $T= 2-5K$ and $P =~1.5 GPa$, a volume change at the phase transition reaches a value almost two order of magnitude higher $-\Delta V/V\approx 2\times10^{-4}$ Ref.\cite{10, 22}.  The question arises what kind of connection exists between these two values and how this great increase of the volume discontinuity is related to the evolution of the phase transition in MnSi?

\begin{figure}[htb]
\includegraphics[width=80mm]{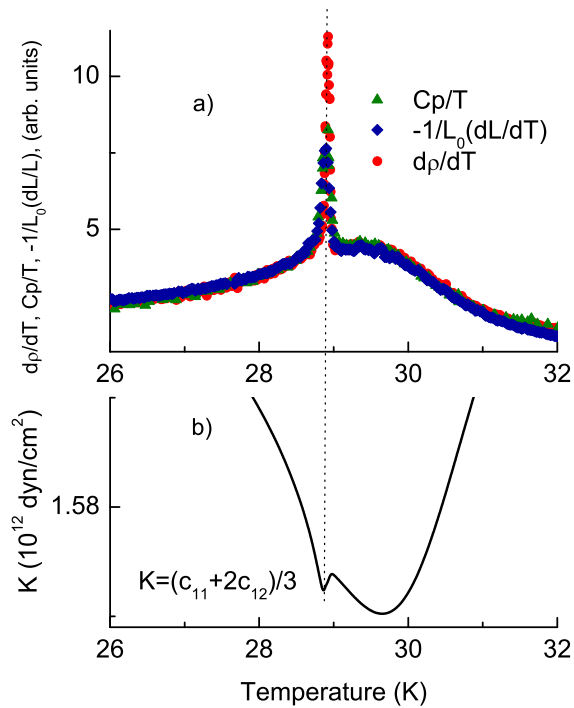}
\caption{\label{fig1} (Color online) a) Reduced heat capacity divided by temperature $C_{p}/T$, linear coefficient of thermal expansion $1/L_{o} (dL/dT)$, temperature derivative of resistivity $d\rho/dT$ and b) bulk modulus $K $of MnSi as functions of temperature in the vicinity of the phase transition.}
 \end{figure}
Let us consider different scenarios for the evolution of the first order transition in MnSi along the phase transition line. (1) The first order phase transition continues with applied pressure down to the lowest temperature, and  a polycritical point does not exist along the phase transition line. This scenario seems hard to reconcile with the above-mentioned values of the volume discontinuities. (2) The first order phase transition, which is driven in the current case to the first order by fluctuations \cite{5}, evolves into a second order one while thermal fluctuations are suppressed  at elevated pressures and lower temperatures. In this case there would be a tricritical point on the phase transition line that excludes any step - like volume discontinuity in the low temperature limit. Quantum fluctuations could be suppressed as well due to an increased effective dimensionality of the quantum system, but this may appear not true in the special case of MnSi \cite{23}.

\begin{figure*}[htb]
\includegraphics[width=180 mm]{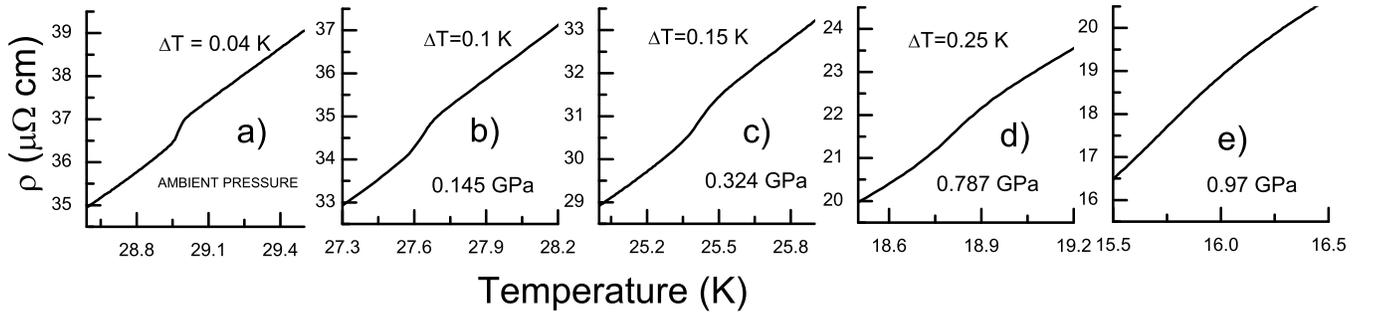}
\caption{\label{fig2} Resistivity change at the phase transition in MnSi . $\Delta T$ is the  temperature width
of the phase transition.}
 \end{figure*}

In an attempt to resolve the issue of the evolution of the phase transition in MnSi, we performed a series of resistivity measurements on MnSi single crystal at high pressures. The reason for this choice lies in the observation of almost perfect scaling between the temperature derivative of resistivity, thermal expansion coefficient and heat capacity \cite{13} (see Fig.\ref{fig1}a). Consequently,  the resistivity also provides information about thermodynamic properties of the phase transition in MnSi. However, one should keep in mind that the magnetic first order transition in MnSi is just a minor feature of the extended anomaly of thermodynamic properties. Figure \ref{fig1}b, displaying the bulk modulus of MnSi as a function of temperature, just emphasizes this fact. 
In the present study we intentionally employed a classical high pressure piston – cylinder technique, analogues to that used in \cite{10, 22}, with silicon liquid as a pressure medium. The single crystal samples were cut from the same batch used in our previous experiments \cite{11,12,13}. The resistivity was measured by a standard four terminal dc technique. Temperature was registered with a Cernox sensor, and pressures were calculated from known coordinates of the phase transition line. 
We do not show a complete set of resistivity curves obtained in the current experiment. Instead, Fig.\ref{fig2} displays only limited parts of the resistivity curves in the vicinity of the phase transition. The form of the resistivity curve at ambient pressure clearly evidences the first order nature of the magnetic transition in MnSi (Fig.\ref{fig2}a). We emphasize the narrow temperature interval of the phase transition $~ 0.04 K $ Fig.\ref{fig2}a), demonstrated by the unique sample used in the current study. Figures \ref{fig2} b,c,d and e show fast degradation of the first order features of the phase transition with pressure. Finally, at a pressure of about 1GPa there no longer is a trace of the first order transition. We will discuss this issue later. 

 \begin{figure}[htb]
\includegraphics[width=80mm]{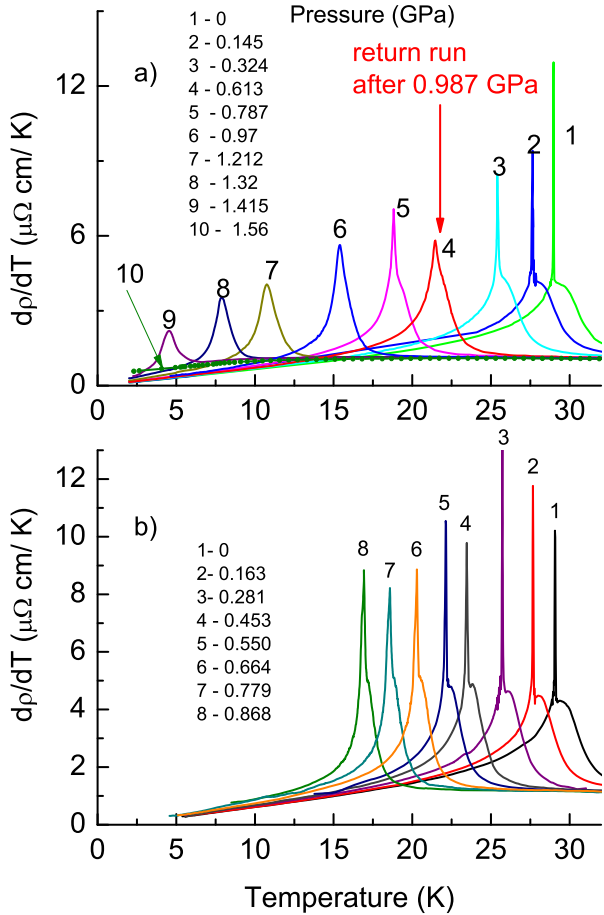}
\caption{\label{fig3} (Color online) Temperature derivatives of resistivity $d\rho/dT$ at the phase transitions in MnSi at different pressures. a) Current data, obtained with a liquid pressure medium and b) data, obtained with helium as a pressure medium \cite{11}. }
 \end{figure}

Figure \ref{fig3}a illustrates the evolution of the temperature derivatives of resistivity $d\rho/dT$ of MnSi along quasi isobars, where they cross the phase transition boundary. The data, obtained in with helium as a pressure medium \cite{11}, are reproduced in Fig.\ref{fig3}b for comparison.
As is seen in Fig. \ref{fig3}a at low pressure $d\rho/dT$ has a now well-known form of sharp peaks with shallow shoulders on the high temperature side of the peak. This sharp peak in the $d\rho/dT$ identifies a first order phase transition (see Fig.\ref{fig2} a for confirmation). On application of pressure the sharp peaks start to smear out and completely disappear between $0.8$ and $1.0 GPa$ in correspondence with Fig. \ref{fig2}. At the first sight it might suggest a change of order of the phase transition; however, there are clear indications of the disruptive influence of non-hydrostatic stresses on the phase transition. Indeed, the return run at a pressure $0.6 GPa$ shows an essential deformation of the form of $d\rho/dT$ after loading the sample to $\sim 1 GPa$. Unfortunately, the experiments with helium were not extended to higher pressures due to technical problems (see Fig. \ref{fig3}b).

\begin{figure}[htb]
\includegraphics[width=80mm]{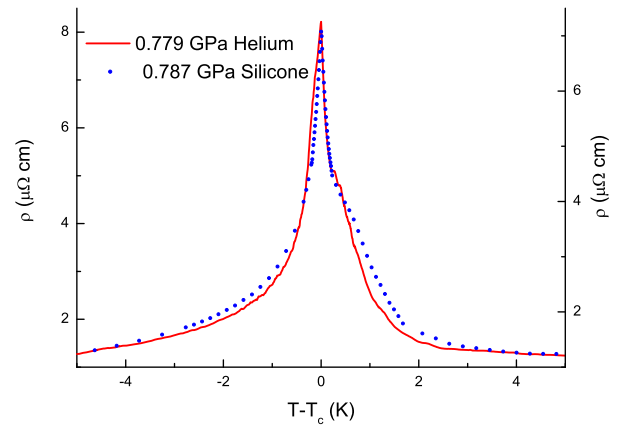}
\caption{\label{fig4} (Color online) Comparison of temperature derivatives of resistivity $d\rho/dT$  of MnSi, obtained in two different pressure media (helium and silicone liquid).}
 \end{figure}

From a comparison Figs. \ref{fig3}a and \ref{fig3}b, it is hard to make a definite conclusion on whether the frozen silicon liquid is inferior to solid helium as a pressure medium. This is emphasized in Fig. \ref{fig4} , where two curves of $d\rho/dT$, obtained from the "silicon" and "helium" experiments  at the same pressure are displayed. Surprisingly, these two curves differ very little. Thus, we are unable to state whether the sharp peaks in $d\rho/dT$ disappeared because of non-hydrostatic stresses or /and some fundamental physics is also involved. Anyway we may conclude that a sharp first order transition in MnSi is not seen and probably can not be seen in experiments at pressures higher than $0.8-1.0 GPa$ with frozen liquids as a pressure medium (see Fig.\ref{fig2}).  We also note that sharp peaks in the thermal expansion coefficient of MnSi was not be seen \cite{10, 22} even at ambient and low pressures due to insufficient resolution.

\begin{figure}[htb]
\includegraphics[width=80mm]{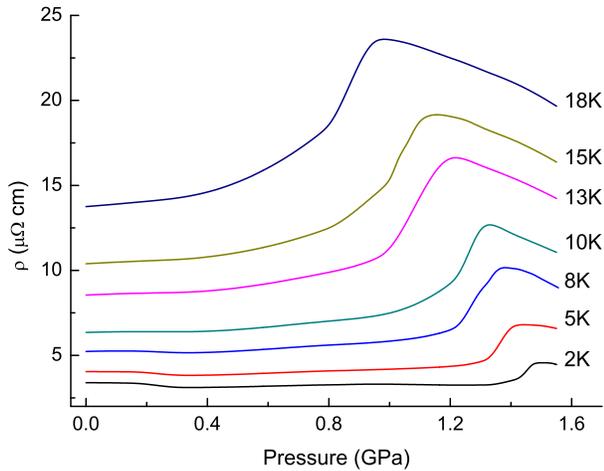}
\caption{\label{fig5} (Color online) Resistivity isotherms of MnSi, showing an evolution of the fluctuation region in the vicinity of the phase transitions.}
\end{figure}

\begin{figure}[htb]
\includegraphics[width=80 mm]{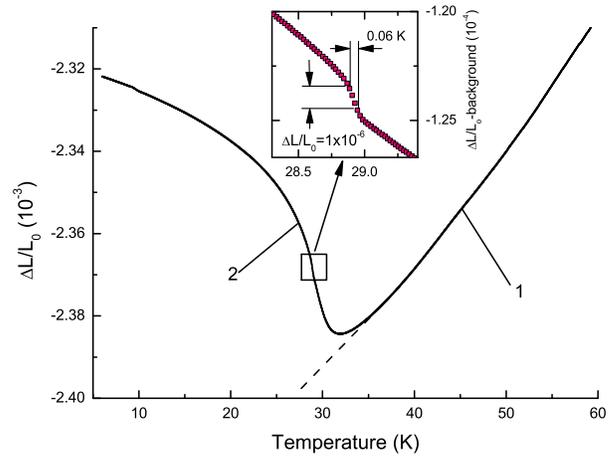}
\caption{\label{fig6} (Color online) Linear thermal expansion of MnSi illustrating a relationship between the volume anomaly and the first order phase transition. 1-"normal " branch with positive thermal expansion, 2-fluctuation branch with negative expansion, $L_{o}$-room temperature value. Modified from \cite{12}.  }
\end{figure}

So what is a nature of the low temperature volume anomaly found in \cite{10, 22}? First, we point out that the magnetic transition in MnSi is accompanied by continuous anomalies in thermodynamic and transport properties induced by the spin fluctuations \cite{24} Fig.\ref{fig1}. At high pressures and lower temperatures these anomalies should become progressively less extended as the fluctuations become frozen-out. Finally, at these pressures and temperatures thermodynamic and transport anomalies may be well defined and look like smeared discontinuities, therefore imitating a first order phase transition. These conclusions are well illustrated by Fig.\ref{fig5} that displays resistivity isotherms crossing the phase transition, constructed from the quasi-isobaric temperature-dependent resistivity measurements.  As is seen in Fig.\ref{fig5}, the resistivity anomaly at the phase transition at low temperature is localized in the vicinity of the phase transition within $~0.1 GPa$ in agreement with results of Ref.\cite{10, 22}. Though one should be cautious trying to draw an exact relation between volume and resistivity in the quantum regime, a volume anomaly by definition is accompanied by a resistivity feature at temperatures above zero. Thus, the claims of experimental observations of a first order phase transition in MnSi at low temperature and high pressures are certainly not valid. Instead, the volume anomaly reported earlier \cite{10, 22} likely is caused by  the spin fluctuations that cannot be classified as a phase transition. This statement is well illustrated in Fig.\ref{fig6}, demonstrating a relationship between the volume anomaly and the phase transition in MnSi at ambient pressure.  As one can see the huge volume anomaly spreading from almost zero to ~35 K practically conceals the tiny first order phase transition (see Fig.\ref{fig6} and the insert ). This transition becomes unobservable at pressures more than $1 GPa$ and only the "squeezed " volume anomaly can be seen at higher pressures, as is reflected in the resistivity behavior (Fig.\ref{fig5}.)

\begin{figure}[htb]
\includegraphics[width=80 mm]{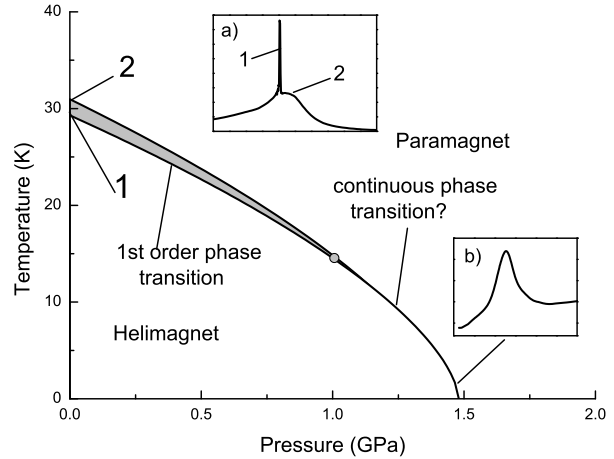}
\caption{\label{fig7} Tentative phase diagram of MnSi at high pressure. The grey region shows the helical fluctuation domain, disappearing at a pressure of $\sim 1 GPa$. The grey circle could be a tricritical point, if the phase transition at pressures $ > 1 GPa $ is truly continuous. a) and b)  illustrate variations of the form $d\rho/dT$ with pressure.}
 \end{figure}

Returning to the phase transition evolution one should be aware of two notable features of the phase transition that disappear with decreasing temperature. They are the sharp peak marking the first order phase transition and the shallow maximum, situated slightly above the critical temperature and pointing to a domain of prominent helical fluctuations\cite{15, 25, 26} (see Fig.\ref{fig1}). In result at low temperature the phase transition in MnSi is manifested by almost symmetric and rather steep maxima in the temperature derivatives of resistivity. 
The current experimental data obviously rule out any strongly first order phase transition at $T\rightarrow 0$ in MnSi, though a weak smeared out first order phase transition may be hidden at the slopes of the maxima. The tentative phase diagram of MnSi is displayed in Fig.\ref{fig7}.

As was emphasized we are unable to distinguish between a first order phase transition smeared by non-hydrostatic stress and a second order phase transition. An evolution of the first order phase transition in MnSi along the phase transition line is illustrated in Fig.\ref{fig2}. In the case of a true second order transition in MnSi at high pressures we should have a tricritical point corresponding to the evolution from a weak first order transition at high temperatures to a second order one at low temperatures. This is quite opposite to former suggestions \cite{7,8,9}.  We place a provisional tricritical point at a position with coordinates 16 K and 1GPa (Fig.\ref{fig6}), where the helical fluctuation domain disappears. We reiterate that the helical fluctuation domain is situated between the helical magneto-ordered phase and the paramagnetic phase of MnSi in a narrow temperature interval.   The simultaneous disappearance of this strongly fluctuating domain and first order features in the phase transition emphasizes a connection between fluctuations and a first order nature of the phase transition in MnSi at ambient pressure. At the same time this situation may prove the inability of quantum fluctuations to influence the order of a phase transition. 
In any case we conclude there should exist true quantum critical phenomena in MnSi at $T\rightarrow 0$ because the tiny first order transition, if it survives to the lowest temperatures, cannot suppress the entire quantum critical region.
In summary, the sharp peaks and shallow shoulders in the temperature derivatives of resistivity of MnSi, which indicate the first order phase transition and the helical fluctuation domain, disappear simultaneously at 16 K and 1GPa. All that may imply an existence of a tricritical point on the phase transition line. On the other hand, it may mean that the first order phase transition, being smeared by non-hydrostatic stresses in a frozen liquid, continues the same way to the lowest temperatures. Note that in fact the imaginable conflict expected early between the volume discontinuity at ambient pressure and the high pressure – low temperature volume anomaly does not exist at all because the mentioned features are related to different physical phenomena. In either case the present experimental data do not support the idea of a strong first order transition in MnSi at $T\rightarrow 0$, which would prevent development of the quantum critical region. On the contrary, one can see evidence of the quantum criticality in MnSi in its non-Fermi liquid resistivity behavior at high pressures \cite{27}. 

This research is supported by the Russian Foundation for Basic Research (12-02-00376-a), Program of the Physics Department of RAS on Strongly Correlated Systems and Program of the Presidium of RAS on Physics of Strongly Compressed Matter.

\end{document}